\pdfoutput=1
\documentclass[preprint,prb,amsmath,superscriptaddress,aps,subeqn,showpacs]{revtex4}

\usepackage{graphicx}

\begin{document}

\title{Intrinsic Electrical Transport Properties of Monolayer Silicene and MoS$_2$ from First Principles}

\author{Xiaodong Li}
\affiliation{Department of Electrical and Computer Engineering, North Carolina State University, Raleigh, NC 27695-7911}

\author{Jeffrey T. Mullen}
\affiliation{Department of Physics, North Carolina State University, Raleigh, NC 27695-8202}

\author{Zhenghe Jin}
\affiliation{Department of Electrical and Computer Engineering, North Carolina State University, Raleigh, NC 27695-7911}

\author{Kostyantyn M. Borysenko}
\affiliation{Department of Physics, Texas Southern University, Houston, TX 77004}

\author{M. Buongiorno Nardelli}
\affiliation{Department of Physics and Department of Chemistry, University of North Texas, Denton, TX 76203} \affiliation{CSMD, Oak Ridge National Laboratory, Oak Ridge, TN 37831}

\author{Ki Wook Kim} \email{kwk@ncsu.edu}
\affiliation{Department of Electrical and Computer Engineering, North Carolina
State University, Raleigh, NC 27695-7911}

\begin{abstract}
The electron-phonon interaction and related transport properties are investigated in monolayer silicene and MoS$_2$ by using a density functional theory calculation combined with a full-band Monte Carlo analysis.  In the case of silicene, the results illustrate that the out-of-plane acoustic phonon mode may play the dominant role unlike its close relative $-$ graphene. The small energy of this phonon mode, originating from the weak sp$^2$ $\pi$ bonding between Si atoms, contributes to the high scattering rate and significant degradation in electron transport.  In MoS$_2$, the longitudinal acoustic phonons show the strongest interaction with electrons.  The key factor in this material appears to be the Q valleys located between the $\Gamma$ and K points in the first Brillouin zone as they introduce additional intervalley scattering.  The analysis also reveals the potential impact of extrinsic screening by other carriers and/or adjacent materials.  Subsequent decrease in the actual scattering rate can be drastic, warranting careful consideration.  Finally, the effective deformation potential constants are extracted for all relevant intrinsic electron-phonon scattering processes in both materials.


\end{abstract}


\maketitle

\section{INTRODUCTION}
Very recently, the attention to low-dimensional materials has expanded beyond the best known example of this kind $-$  graphene.~\cite{Novoselov2005MoS2,Wang2012_2D,Coleman2011,Sahin2009} In particular, silicene~\cite{Kara2012,Cahangirov2009,Vogt2012} and molybdenum disulfide~\cite{Mak2010,Radisavljevic2011,Kim2012,Kong2012} have gained much interest due to their  unique properties in electronics, optoelectronics and magnetics. Silicene is expected to share certain superior properties of graphene due to its structural similarity  and the close position in the periodic table.  More importantly, it is compatible with the current silicon-based technology and can be grown on a number of different substrates.~\cite{Chen2012_Silicene,Feng2012,Fleurence2012}  On the other hand, atomically thin MoS$_2$ is a semiconductor with a finite band gap that ranges from approximately 1.3~eV to 1.9~eV depending on the thickness.~\cite{Mak2010} It has been used as the channel material in field effect transistors with promising results.~\cite{Radisavljevic2011,Lembke2012} In addition, monolayer MoS$_2$ offers the possibilities of interesting spin and valley physics utilizing the strong spin-orbit coupling.~\cite{Xiao2012,Cao2012,Zeng2012,Mak2012}

Characterization of electronic transport, particularly the intrinsic  properties, is critical for assessing and understanding the potential significance of a material.  In the case of silicene, many of the crucial parameters are presently unknown due to the brief history of this material.  In comparison, notable advances have been made in MoS$_2$ lately.  Experimental investigation of transistor characteristics
deduced the channel mobilities ranging from $\sim$200~cm$^2$/Vs to $\sim$1000~cm$^2$/Vs at room temperature,~\cite{Radisavljevic2011,Lembke2012}  while a theoretical study estimated an intrinsic phonon-limited value of $\sim$410 cm$^2$/Vs based on a first-principles calculation of electron-phonon interaction.~\cite{Kaasbjerg2012} However, questions remain regarding the intrinsic electron transport in MoS$_2$.  For instance, those extracted from transistor current-voltage (I-V) measurements are indirect accounts and can be strongly affected by extrinsic factors.  Similarly, the latter work~\cite{Kaasbjerg2012} includes only the electronic states in the conduction band minima at the K points in the first Brillouin zone (FBZ); the impact of Q valleys located along the ${\Gamma}$-K symmetry directions (which correspond to the minima of bulk MoS$_2$) were not considered.  A detailed investigation is clearly called for.

In this paper, we examine intrinsic transport properties of monolayer silicene and MoS$_2$ by taking advantage of first-principles analysis and full-band Monte Carlo simulation.  Along with the electronic band structure, the phonon spectra and electron-phonon coupling matrix elements are calculated for all phonon branches within the density functional theory (DFT) formalism.~\cite{Baroni2001,Borysenko2010}
The obtained electron scattering rates are subsequently used in the Boltzmann transport equation to compute the intrinsic velocity-field characteristics  with a full-band Monte Carlo treatment.  The calculation results are compared with the available data in the literature and the key factors affecting electron transport in these materials elucidated.
The investigation also provides the effective deformation potential constants extracted from the first-principles results.


\section{THEORETICAL MODEL}

Both monolayer silicene and MoS$_2$ are hexagonal crystals.  To account for their delicate atomic structures accurately, the calculations are performed in the DFT framework, as it is implemented in the QUANTUM-ESPRESSO package,~\citep{Giannozzi2009} using ultrasoft pseudopotentials.  A minimum of 35 Ry is used for the energy cut-off in the plane wave expansion along with the charge truncation
$\sim$15 times larger.  The generalized gradient approximation is used for the exchange-correlation potential for silicene, while the local density approximation is adopted for MoS$_2$.  The momentum space is sampled on a 36$\times$36$\times$1 Monkhorst-Pack grid with no offset (silicene) or on a 18$\times$18$\times$1 grid (MoS$_2$) for electronic band calculation.  The simulated cells are optimized in both cases until the atomic forces decrease to values less than 0.015~eV/{\AA}.

Each phonon is treated as a perturbation of the self-consistent potential, created
by all electrons and ions of the system, within the linear response [i.e., density functional perturbation theory (DFPT)].~\cite{Baroni2001}  The calculation of the potential change due to this perturbation gives the value of the electron-phonon
interaction matrix element:~\cite{Borysenko2010}
\begin{equation}\label{matrix}
\textsf{g}_{\mathbf{q},\mathbf{k}}^{(i,j)v}=\sqrt{\frac{\hbar}{2M\omega_{v, \mathbf{q}}}}\langle j,\mathbf{k}+\mathbf{q}\vert \Delta V_{\mathbf{q},\mathrm{SCF}}^{v}\vert i,\mathbf{k}\rangle,
\end{equation}
where
$|i, \mathbf{k} \rangle$ is the Bloch electron eigenstate with the wavevector $\mathbf{k}$, band index $i$, and energy $E_{i, \mathbf{k}}$; $\Delta V_{\mathbf{q},\mathrm{SCF}}^{v}$ is the derivative
of the self-consistent Kohn-Sham potential~\cite{Baroni2001} with respect to atomic displacement associated with the phonon from branch $v$ with the wavevector $\mathbf{q}$ and frequency
$\omega_{v, \mathbf{q}}$; and M is the atomic mass.   Numerical calculations of lattice dynamics are conducted on a 18$\times$18$\times$1 Monkhorst-Pack grid.  Indices $i,j$  are dropped hereinafter as only the first (lowest) conduction band is considered.

With the electron-phonon interaction matrix from the first-principles calculation, the scattering rate of an electron at state $| \mathbf{k} \rangle$  can be obtained by using Fermi's golden rule,
\begin{align}\label{Fermi_Golden}
\frac{1}{\tau_{\mathbf{k}}} =\frac{2\pi}{\hbar}\sum_{\mathbf{q},v}\vert\textsf{g}_{\mathbf{q},\mathbf{k}}^v\vert^2
[N_{v,\mathbf{q}}\delta(E_{\mathbf{k}+\mathbf{q}}-\hbar\omega_{v,\mathbf{q}}-E_{\mathbf{k}})
+(N_{v,\mathbf{q}}+1)\delta(E_{\mathbf{k}-\mathbf{q}}+\hbar\omega_{v,\mathbf{q}}-E_{\mathbf{k}})],
\end{align}
where $N_{v,\mathbf{q}}=[\exp({\hbar\omega_{v,\mathbf{q}}}/{k_B T})+1]^{-1}$ is the phonon occupation number, $k_B$ the Boltzmann constant, and $T$ the temperature.  As we are interested in the intrinsic scattering probability that is not limited to a specific carrier distribution (and thus, the Fermi level), our formulation assumes that all final electronic states are available for transition (i.e., nondegenerate) in the bands under consideration.

For transport properties, a Monte Carlo approach with full-band treatment is adopted.  All of the details described above, including the wavevector ($\mathbf{k}$,$\mathbf{q}$)  dependence of the scattering matrix elements [e.g., Eq.~(\ref{matrix})], are accounted for.  This allows solution of the Boltzmann transport equation beyond the conventional relaxation time approximation.

\section{RESULTS AND DISCUSSION}
\subsection{Monolayer Silicene}
Earlier first-principles studies have shown that the stable structure for monolayer silicene has a low-buckled configuration.~\cite{Sahin2009,Cahangirov2009}  While planar and high-buckled cases lead to imaginary phonon frequencies around the $\Gamma$ point indicating an unstable structure, the low-buckled construction provides well separated phonon branches and positive frequencies. The origin of buckled geometry in silicene is the weakened $\pi$ bonding of the electrons in the outer shell. Compared with graphene, which has very strong $\pi$ bonding and planar geometry, the interatomic bonding distance is much larger in silicene, which decreases the overlap of $p_z$ orbitals and dehybridizes the $sp^2$ states. Accordingly, the planar geometry cannot be maintained. In our analysis, the lattice constant $a$ is optimized to be 3.87 {\AA}, with the buckling distance of 0.44 {\AA}, in good agreement with Ref.~\onlinecite{Sahin2009}.

\begin{figure}
\centering
\includegraphics[width=8.0cm,height=4.5cm]{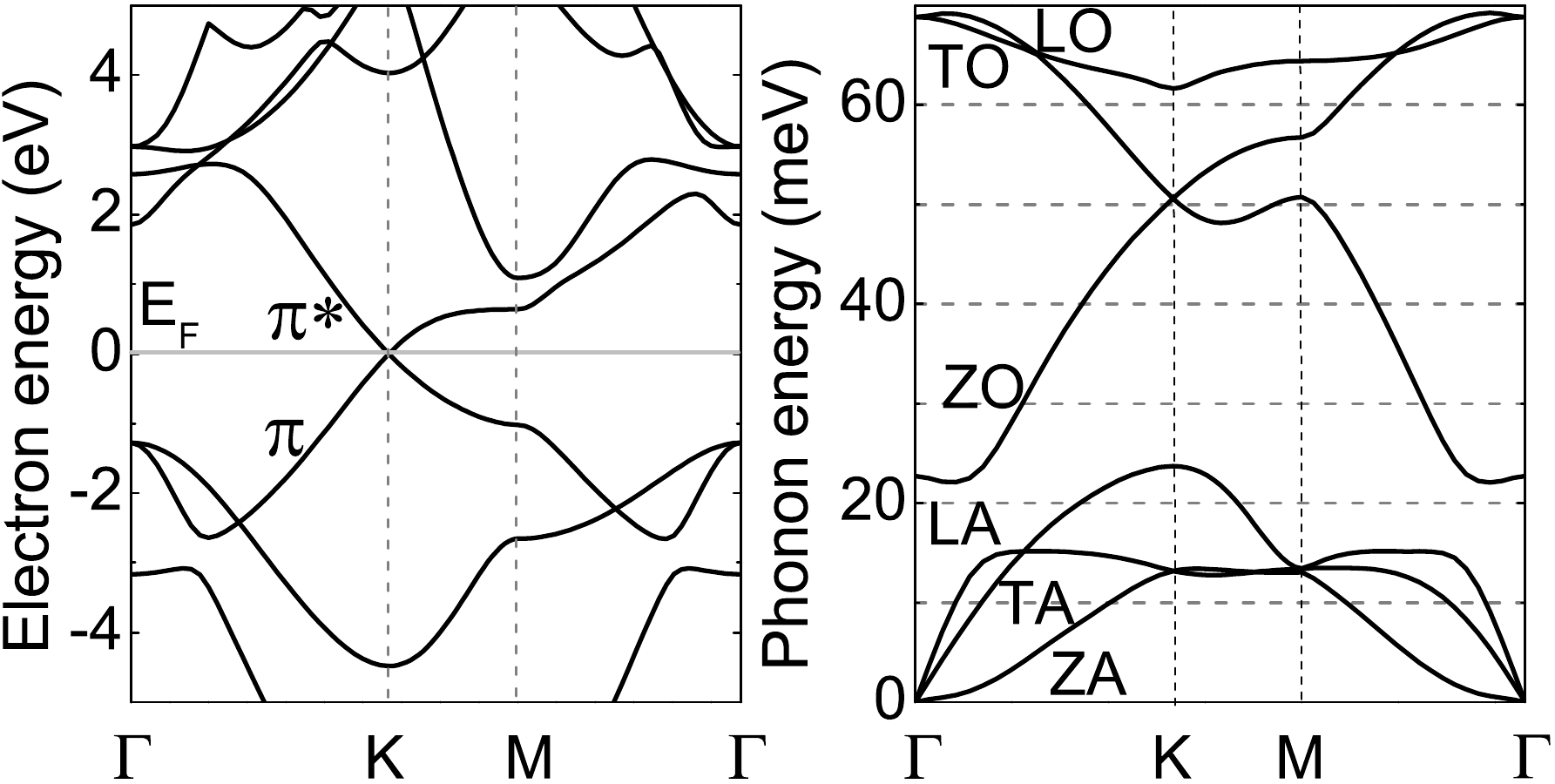}
\caption{Electronic and phononic band structures of monolayer silicene along the symmetry directions in the FBZ.   The Dirac point serves as the reference of energy scale for electrons.}
\label{Silicene_band_ph}
\end{figure}

Figure~\ref{Silicene_band_ph} shows the outcome of electronic and phononic band calculation in monolayer silicene.  The  Fermi velocity extracted from the Dirac cone  is around $5.8 \times 10^7$~cm/s that is roughly one half of that in graphene [see bands $\pi$ and $\pi^*$ in Fig.~\ref{Silicene_band_ph}(a)]. While this result is in agreement with other theoretical predictions,~\cite{Cheng2011,Drummond2012} a value as high as $1 \times 10^8$~cm/s was also claimed in the literature.~\cite{Cahangirov2009} As for phonons in Fig.~\ref{Silicene_band_ph}(b), six branches are identified with two atoms per unit cell. The transverse acoustic (TA) and longitudinal acoustic (LA) phonon dispersion relations are well approximated by sound velocities in the long-wavelength limit; $v_\mathrm{TA}=5.4\times 10^5$~cm/s and $v_\mathrm{LA}=8.8\times 10^5$~cm/s. Although the out-of-plane acoustic (ZA) phonon exhibits an approximate q$^2$ dependence near the center of the Brillouin zone, its sound velocity can also be estimated; $v_\mathrm{ZA}=6.3\times 10^4$~cm/s. An interesting point to note in the phonon dispersion is the discontinuities in the frequency derivative of the highest optical branch that, similar to graphene, appear at the high symmetry points, $\Gamma$ and K. These discontinuities are referred to as Kohn anomalies,~\cite{Piscanec2004,Cheng2011} induced by unusual screening of lattice vibrations by conduction electrons. Sharp cusps typically indicate strong electron-phonon coupling.  Calculated phonon energies at the $\Gamma$, M, and K points in the FBZ are summarized in Table~\ref{table_ph_Silicene}.

\begin{figure}
\centering
\includegraphics[width=8.0cm]{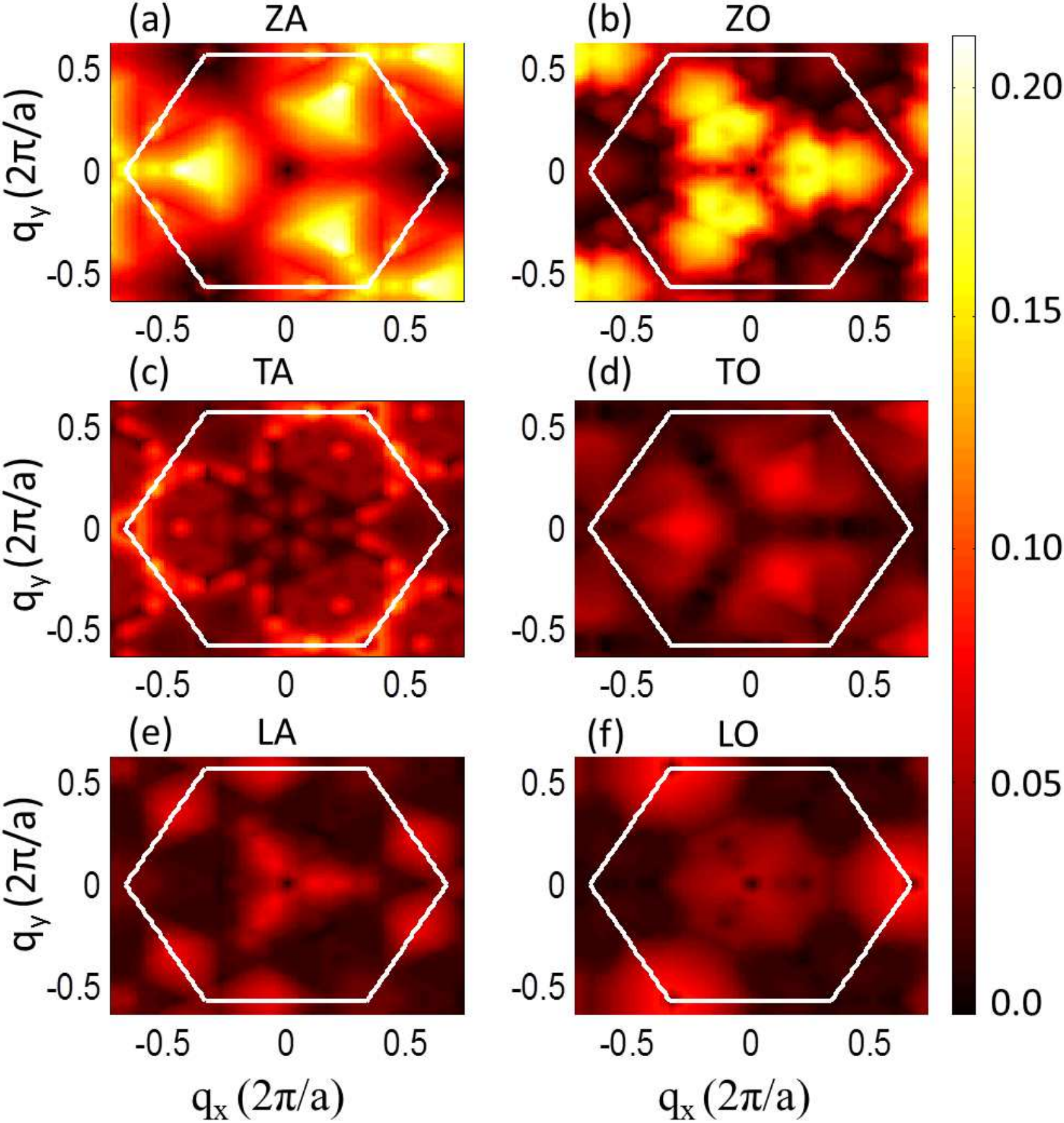}
\caption{(Color online) Electron-phonon interaction matrix elements $\vert\textsf{g}_{\mathbf{k}+\mathbf{q},\mathbf{k}}^v\vert$ (in units of eV) from the DFPT calculation in silicene for $\mathbf{k}$ at the conduction-band minimum K point [i.e., $(4\pi/3a,0)$] as a function of phonon wavevector $\mathbf{q}$ for all six modes $v$.}
\label{Silicene_matrix}
\end{figure}

The electron-phonon interaction matrix elements obtained for the electron at the Dirac point [i.e., $\mathbf{k} = (4\pi/3a,0)$] are plotted in Fig.~\ref{Silicene_matrix} as a function of phonon wavevector $\mathbf{q}$.  Kohn anomalies, illustrated by the three peaks at three equivalent K points in the transverse optical (TO) mode and another at the zone center for the longitudinal optical (LO) branch, are not as distinct as those observed in graphene.~\cite{Borysenko2010} Overall, coupling of optical phonons with electrons appears to be relatively weak. In comparison, the acoustic phonons show much stronger interaction.  Particularly striking is the large strength of ZA phonon coupling, unlike in graphene.  Due to the buckled geometry (originating from the weak $\pi$ bonding mentioned earlier), silicene does not maintain certain key characteristics of ideal planar lattice, especially the reflection symmetry with respect to the atomic plane.  As such, the symmetry consideration, in which only the in-plane phonons can couple linearly to two-dimensional (2D) electrons,~\cite{Manes2007} no longer applies.  An increased role of ZA phonons is clearly expected.

\begin{figure}
\centering
\includegraphics[width=8.0cm]{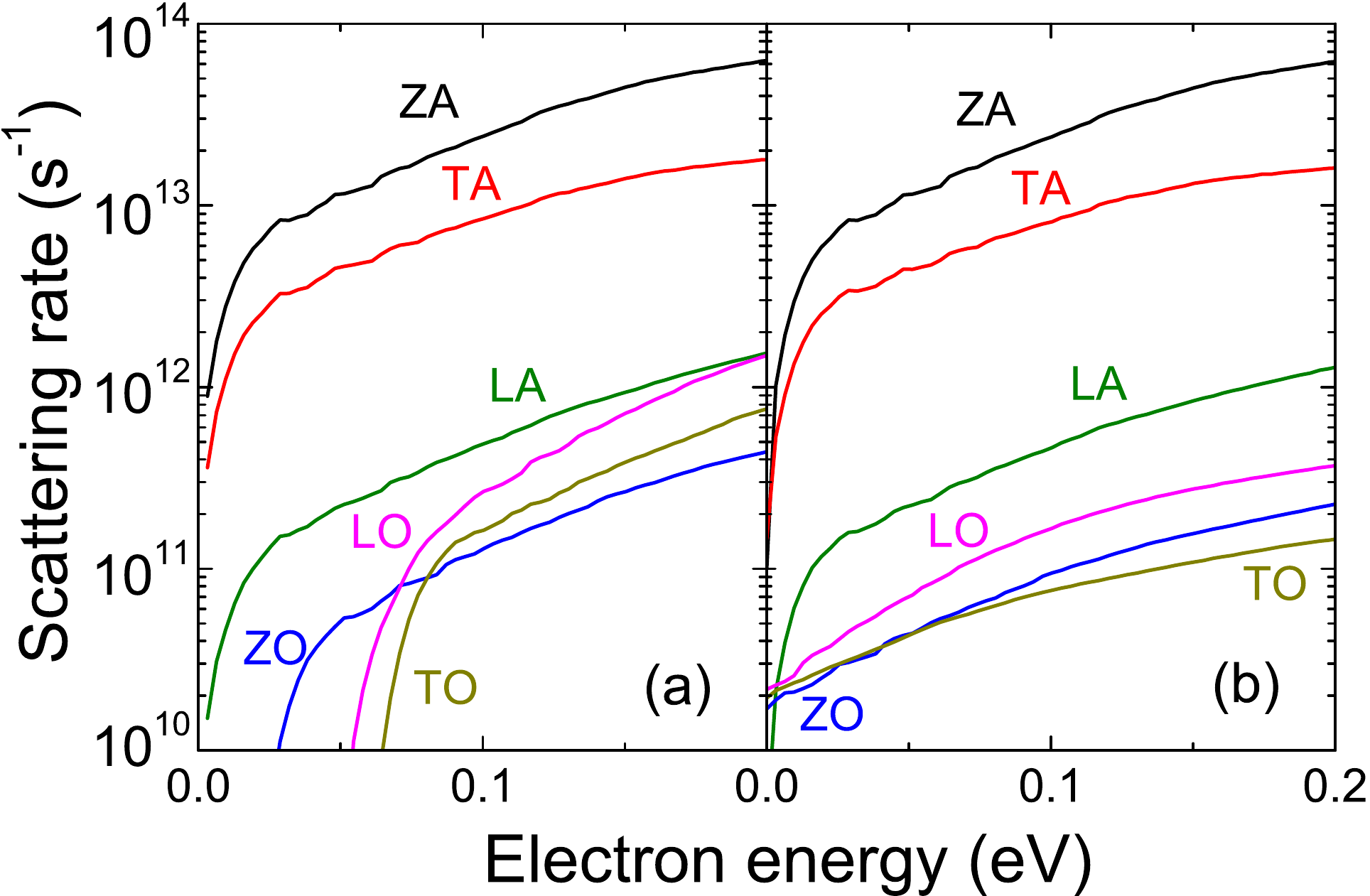}
\caption{(Color online) Electron scattering rates in silicene via (a) emission and (b) absorption of phonons calculated at room temperature. The electron wavevector $\mathbf{k}$ is assumed to be along the K-$\Gamma$ axis.}
\label{Silicene_SR}
\end{figure}

The scattering rates calculated at room temperature ($T = 300$ K) are shown in Fig.~\ref{Silicene_SR}. The result is plotted specifically for electrons with wavevector $\mathbf{k}$ along the K-$\Gamma$ direction. Since the integration in Eq.~(\ref{Fermi_Golden}) is over the entire FBZ, both intravalley ($\mathrm{K}\rightarrow\mathrm{K}$) and intervalley ($\mathrm{K}\rightarrow\mathrm{K}^\prime$) transition events are included. As the interaction matrix elements illustrated above suggest, acoustic phonons have much larger scattering rates than optical modes.  Specifically, the ZA branch provides the dominant contribution, which can be attributed to the observed large coupling strength as well as the small phonon energy near the zone center (i.e., a large occupation number $N_{\mathrm{ZA},\mathbf{q}}$). This also indicates that the scattering rates are very sensitive to the phonon energies (or equivalently the value of $v_\mathrm{ZA}$).   Since an accurate description of ZA dispersion in the long wavelength limit requires a well converged calculation with a sufficiently dense grid, care must be taken when evaluating accuracy of the data in the literature.~\cite{Drummond2012,Cahangirov2009}

\begin{figure}
\centering
\includegraphics[width=8.0cm]{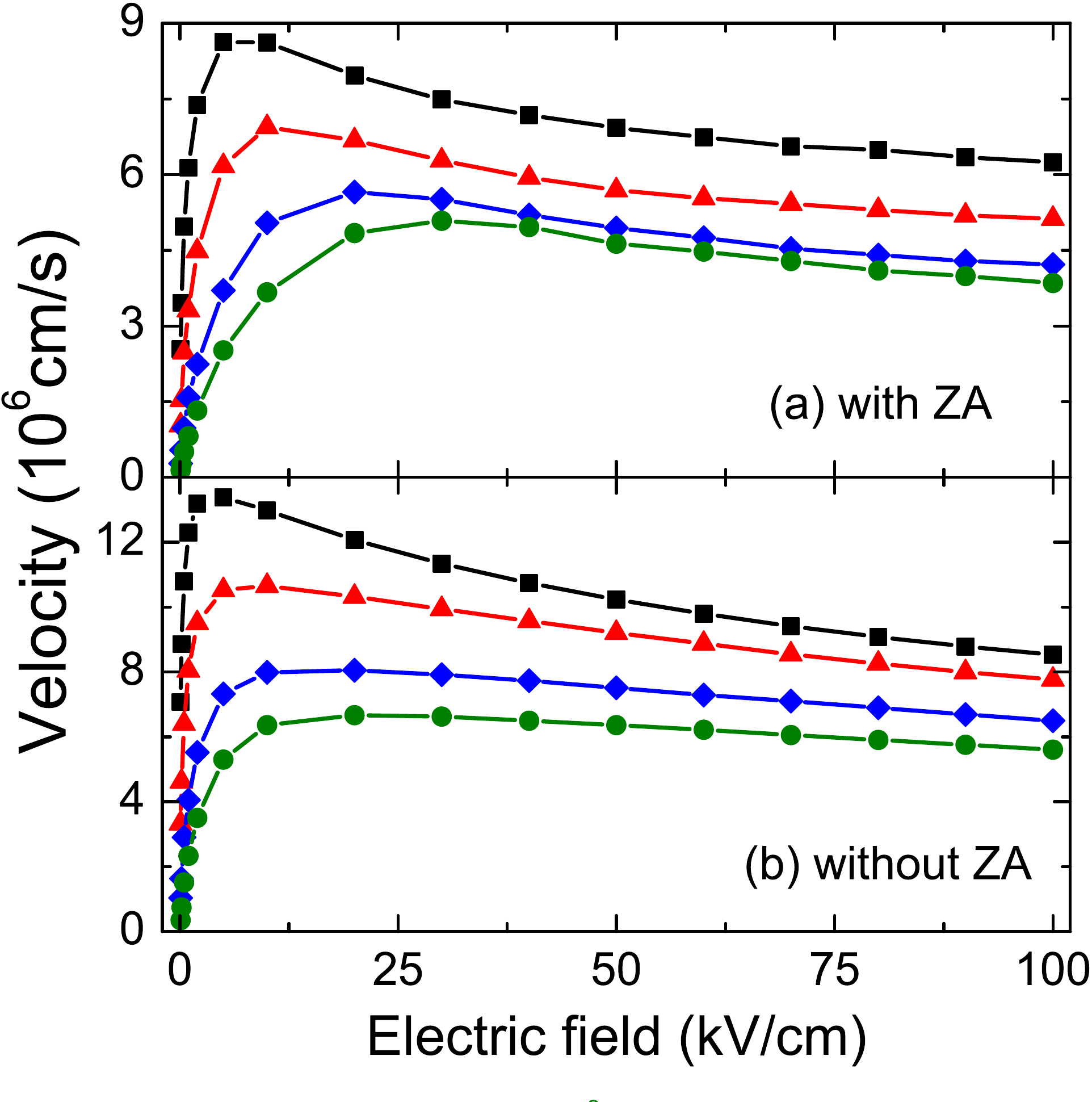}
\caption{(Color online) Drift velocity versus electric field in monolayer silicene obtained from a Monte Carlo simulation at different temperatures: 50 K (square), 100 K (triangle), 200 K (diamond), and 300 K (circle).  The results in (a) consider the scattering by ZA phonons, while those in (b) do not.}
\label{Silicene_vE}
\end{figure}

Figure~\ref{Silicene_vE}(a) provides the drift velocity versus electric field at different temperatures obtained by full-band Monte Carlo simulations.   The intrinsic mobility estimated from the figure is approximately 1200~cm$^2$/Vs and the saturation velocity (defined at 100 kV/cm) $3.9\times 10^6$~cm/s at 300 K. When the temperature decreases, both the mobility and the saturation velocity enhance due to the suppression of phonon excitation; the respective values at 50~K are 3.0$\times$10$^4$~cm$^2$/Vs and $6.2\times 10^6$~cm/s. The drift velocities show a slight negative slope at high fields that becomes more pronounced at low temperatures. This phenomenon (i.e., the negative differential resistance) can be explained, at least in part, by the nonlinear band dispersion at high electron energies as in graphene.~\cite{Li2011}

The calculation results discussed above demonstrate the intrinsic properties of silicene.  When this material is synthesized or placed on a substrate, however, additional scattering sources such as surface polar phonons and impurities must be considered, which could degrade the performance further. A topic that may need additional attention is the role of ZA phonons in the presence of a supporting material.  As  recent measurement of graphene in-plane thermal conductivity attests,~\cite{Cai2010} even a weak binding between a 2D crystal and the substrate could dampen the ZA vibrations substantially.  Moreover, it is reasonable to anticipate that the extent of this suppression would be dependent on the detailed interaction between two materials.  Since ZA phonons provide the dominant role in the electron-phonon interaction in silicene, it (i.e., the damped vibration) could actually lead to sizable reduction in the scattering rate.  To gauge the impact, transport characteristics are also studied without the ZA scattering. As shown in Fig.~\ref{Silicene_vE}(b), the mobility experiences an increase of greater than threefold (3900 cm/s), while the saturation velocity goes up more modestly ($5.6 \times 10^6$~cm/s).  This estimate may be considered an upper limit for silicene on a substrate.

\subsection{Monolayer MoS$_2$}
\begin{figure}
\centering
\includegraphics[width=8.0cm,height=4.5cm]{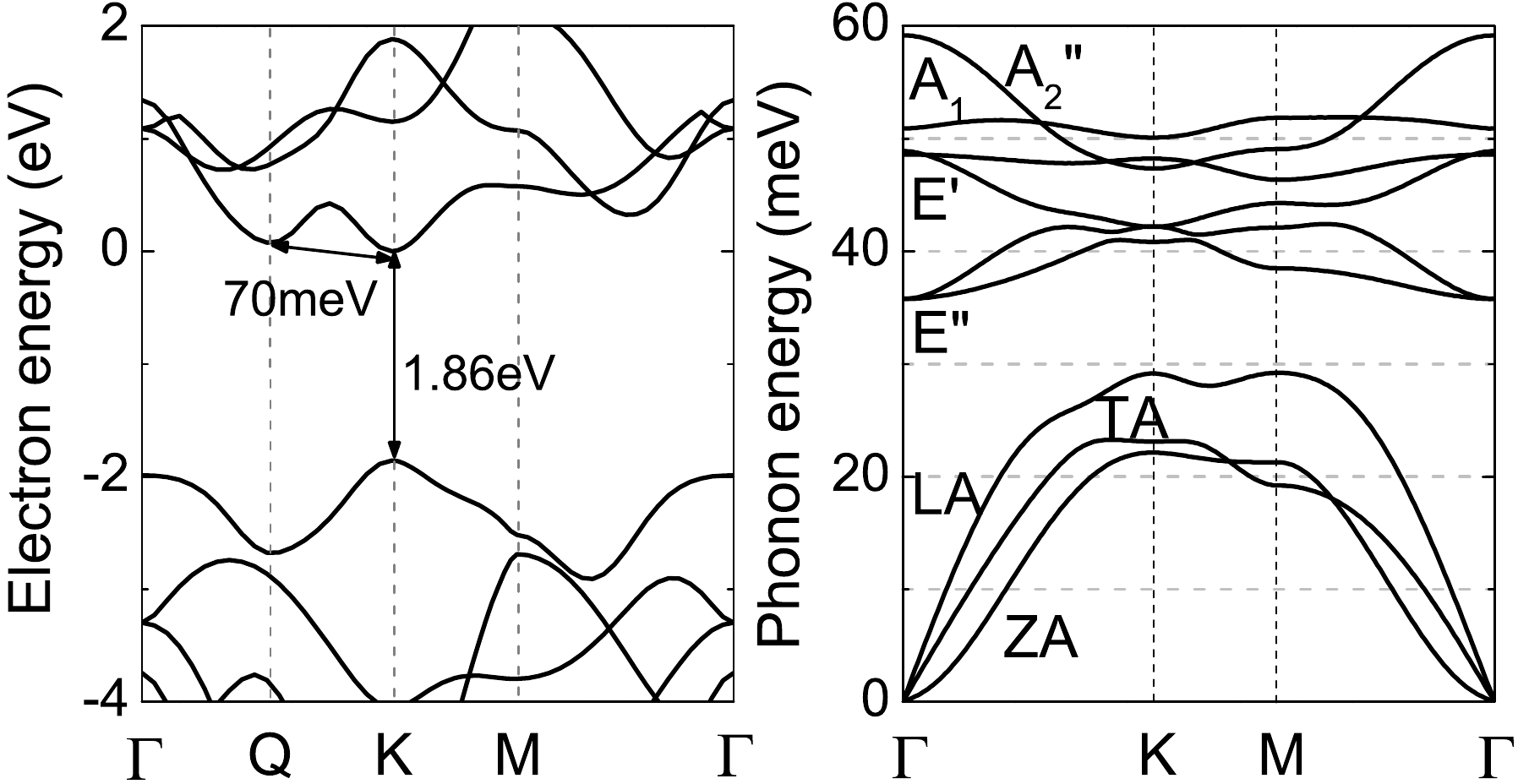}
\caption{Electronic and phononic band structures of monolayer MoS$_2$ along the symmetry directions in the FBZ.  The conduction-band minimum at the K point serves as the reference of energy scale for electrons.}
\label{MoS2_band_ph}
\end{figure}

In the present DFT calculation for monolayer MoS$_2$, the optimized lattice constant is 3.13~{\AA}, consistent with other theoretical studies.~\cite{Kadantsev2012,Kaasbjerg2012,Molina2011} Furthermore, this value is in good agreement with $3.15$~{\AA} determined experimentally in bulk MoS$_2$.~\cite{Wakabayhi1975} The resulting electronic and phononic band dispersion is depicted in Fig.~\ref{MoS2_band_ph}.  As shown, monolayer MoS$_2$ is a semiconductor with a direct gap of 1.86 eV at the K point $-$ a number within a few percent from a recent measurement of 1.9 eV.~\cite{Mak2010}  Our calculation also predicts the presence of second energy minima only about 70~meV  higher.  These so-called Q valleys are located along the $\Gamma$-K axes at approximately the half-way points [e.g., Q=$(0.34\times{2\pi}/{a},0)$ vs.\ K=$({4\pi}/{3a},0)$].  At present, the energy separation $E_\mathrm{QK}$ between the K and Q valleys is unsettled with the estimates ranging from 50 meV to 200 meV.~\cite{Kadantsev2012,Kaasbjerg2012} Since this is a crucial parameter for electron transport, a more extended discussion is provided later in the paper in relation to intervalley scattering. The band dispersion relations around the energy minima are nearly quadratic and can be well described by the effective mass approximation. For the K valleys (i.e., the energy minima at the equivalent K points), the extracted longitudinal and transverse effective masses are almost identical, $m^l_\mathrm{K}=m^t_\mathrm{K}=0.50 m_0$. On the other hand, the Q valleys yield $m^l_\mathrm{Q}=0.62 m_0$, $m^t_\mathrm{Q}=1.0 m_0$ with the longitudinal direction defined along the $\Gamma$-K axis. $m_0$ denotes the electron rest mass.

Monolayer MoS$_2$ has the symmetry of the point group $D_{3h}$, with nine branches of phonons. The irreducible representations associated with each phonon mode, together with the polarization (longitudinal or transverse), help denote all of the vibrational modes,~\cite{Molina2011,Lee2010} as plotted in Fig.~\ref{MoS2_band_ph}. The E$^{\prime\prime}$ modes are degenerate at the $\Gamma$ point. These two modes are the in-plane optical vibrations, with two S atoms moving out of phase and Mo atom static. The E$^\prime$ modes are polar LO and TO phonons, with Mo atom and two S atoms moving out of phase. Due to the coupling with the macroscopic electric field, there is LO-TO separation at the $\Gamma$ point, which slightly lifts the LO(E$^\prime$) mode upward on energy scale. A non-analytical part is added to the dynamic matrix resulting in a small splitting of about 0.3 meV [not visible due to the energy scale of Fig.~\ref{MoS2_band_ph}(b)].  The A$_1$ and A$_2^{\prime\prime}$ branches are two out-of-plane optical phonon modes.  A$_1$ is also referred to as the homopolar mode, with two S atoms moving out of phase and Mo atom static.  In the A$_2^{\prime\prime}$ mode, Mo atom and two S atoms vibrate out of phase.  The three lowest branches are LA, TA, and ZA modes, with sound velocities of $v_\mathrm{LA}=6.6\times 10^5$~cm/s, $v_\mathrm{TA}=4.3\times 10^5$~cm/s. The phonon energies at different symmetric points are summarized in Table~\ref{table_ph_MoS2}.

\begin{figure}
\centering
\includegraphics[width=8.0cm]{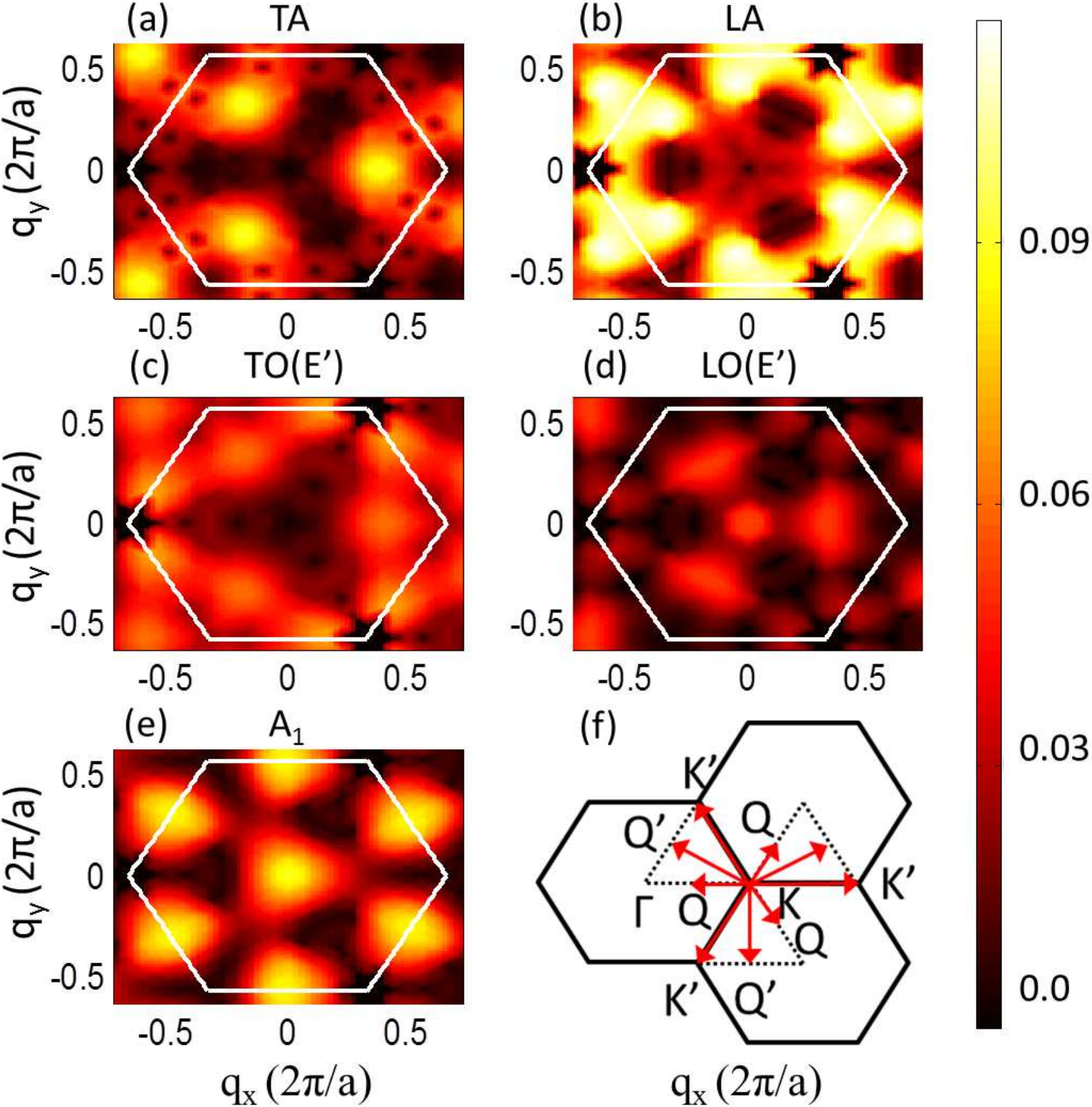}
\caption{(Color online) (a)-(e) Electron-phonon interaction matrix elements $\vert\textsf{g}_{\mathbf{k}+\mathbf{q},\mathbf{k}}^v\vert$ (in units of eV) from the DFPT calculation in MoS$_2$ for $\mathbf{k}$ at the conduction-band minimum K point [i.e., $(4\pi/3a,0)$] as a function of phonon wavevector $\mathbf{q}$.  Only the branches with significant contribution are plotted; i.e., TA, LA, TO(E$^\prime$), LO(E$^\prime$), and A$_1$ (or homopolar) modes. (f) Schematic illustration of {\em intervalley} scattering for electrons in the K valley.}
\label{MoS2_Matrix}
\end{figure}

\begin{figure}
\centering
\includegraphics[width=8.0cm]{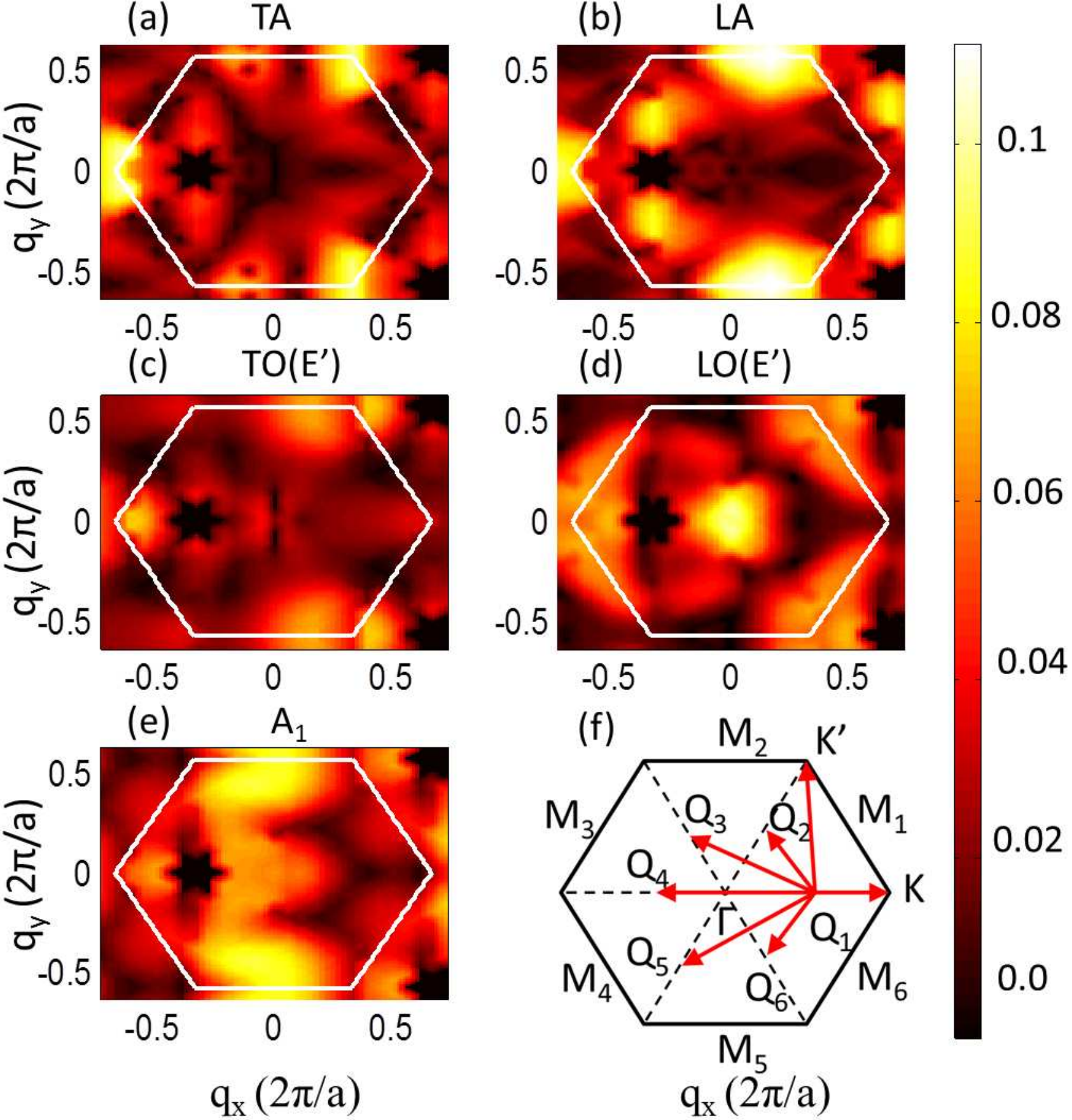}
\caption{(Color online) (a)-(e) Electron-phonon interaction matrix elements $\vert\textsf{g}_{\mathbf{k}+\mathbf{q},\mathbf{k}}^v\vert$ (in units of eV) from the DFPT calculation in MoS$_2$ for $\mathbf{k}$ at the Q point [i.e., $\mathrm{Q_1} \approx(2\pi/3a,0)$] as a function of phonon wavevector $\mathbf{q}$. Only the branches with significant contribution are plotted; i.e., TA, LA, TO(E$^\prime$), LO(E$^\prime$), and A$_1$ (or homopolar) modes. (f) Schematic illustration of {\em intervalley} scattering for electrons in the Q valleys.}
\label{MoS2_MatrixQ}
\end{figure}

Figures~\ref{MoS2_Matrix} and \ref{MoS2_MatrixQ} show the electron-phonon interaction matrix elements for the initial electron state at $\mathbf{k}= \mathrm{K}$ [=$({4\pi}/{3a},0)$] and $\mathbf{k}=\mathrm{Q}$ [$\approx$ $({2\pi}/{3a},0)$], respectively, for TA, LA, TO(E$^\prime$), LO(E$^\prime$), and A$_1$ (or homopolar) phonon modes. The contribution from the remaining four branches is found to be negligible due to the weak coupling. The matrix elements for $\mathbf{k}=\mathrm{K}$ demonstrate a threefold rotational symmetry (i.e., 120$^\circ$), while those of $\mathbf{k} = \mathrm{Q}$ show the reflection symmetry with respect to the $q_x$ axis.  As expected, the LO(E$^\prime$) phonons near the $\Gamma$ point possess the characteristics of Fr\"{o}hlich coupling through the induced macroscopic electric field typical of polar materials. Since the relative electronic potential is not periodic in the long wavelength limit,~\cite{Baroni2001} DFPT does not yield a correct value to the electron-phonon interaction matrix. For an approximation, the matrix element of LO(E$^\prime$) is interpolated at ${\Gamma}$ by using the values from the nearby $q$ points.  This [i.e., LO(E$^\prime$)] and A$_1$ are the only two modes that have non-zero scattering matrix as $\mathbf{q} \rightarrow 0$ (intravalley scattering); in the other three branches, the matrix elements only have first-order components, $|\textsf{g}_{\mathbf{q},\mathbf{k}}|\sim \mathbf{q}$, leading to $|\textsf{g}_{\mathbf{q} \rightarrow 0,\mathbf{k}}|\rightarrow0$.  With regard to intervalley scattering that requires large $q$ phonons, a number of different transition processes are possible as shown in Figs.~\ref{MoS2_Matrix}(f) and \ref{MoS2_MatrixQ}(f).  For instance, Fig.~\ref{MoS2_Matrix} indicates strong electron-phonon interaction at the symmetry points M in the phonon momentum space (denoted as
$\mathbf{q}= \mathrm{M}$ for simplicity) for all modes except LO(E$^\prime$); these phonons can induce electron transition from K to Q$^\prime$ valleys.  Another example is
the phonons at $\mathbf{q}=\mathrm{K^\prime}$ for all five modes in Fig.~\ref{MoS2_MatrixQ}, which can be associated with electron scattering from $\mathrm{Q_1}$ to $\mathrm{Q_4}$.

\begin{figure}
\centering
\includegraphics[width=8.0cm]{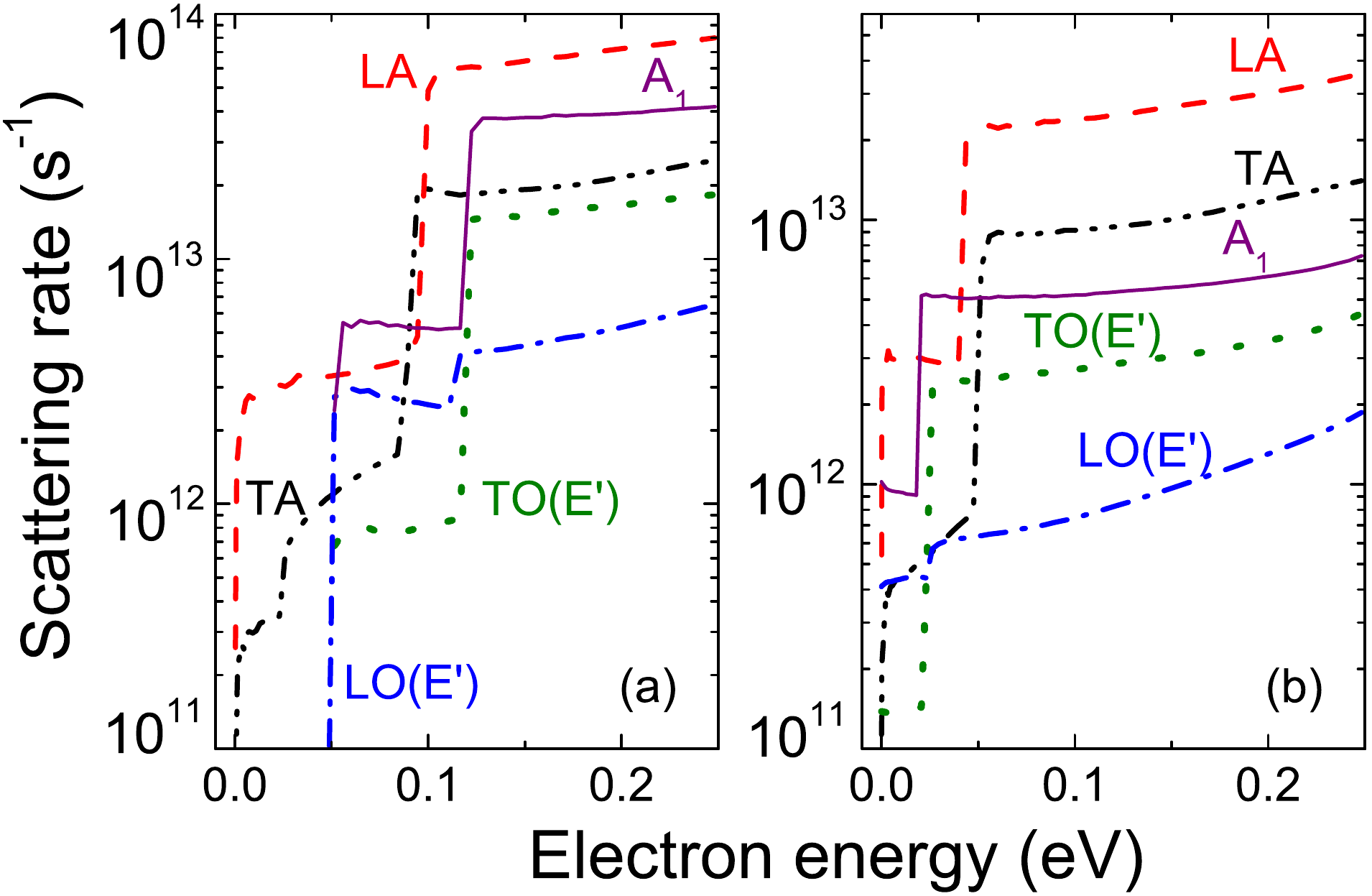}
\caption{ (Color online) Scattering rates of K-valley electrons in MoS$_2$ via (a) emission and (b) absorption of phonons  calculated at room temperature. The electron wavevector $\mathbf{k}$ is assumed to be along the K-$\Gamma$ axis.}
\label{MoS2_SR_K}
\end{figure}

\begin{figure}
\centering
\includegraphics[width=8.0cm]{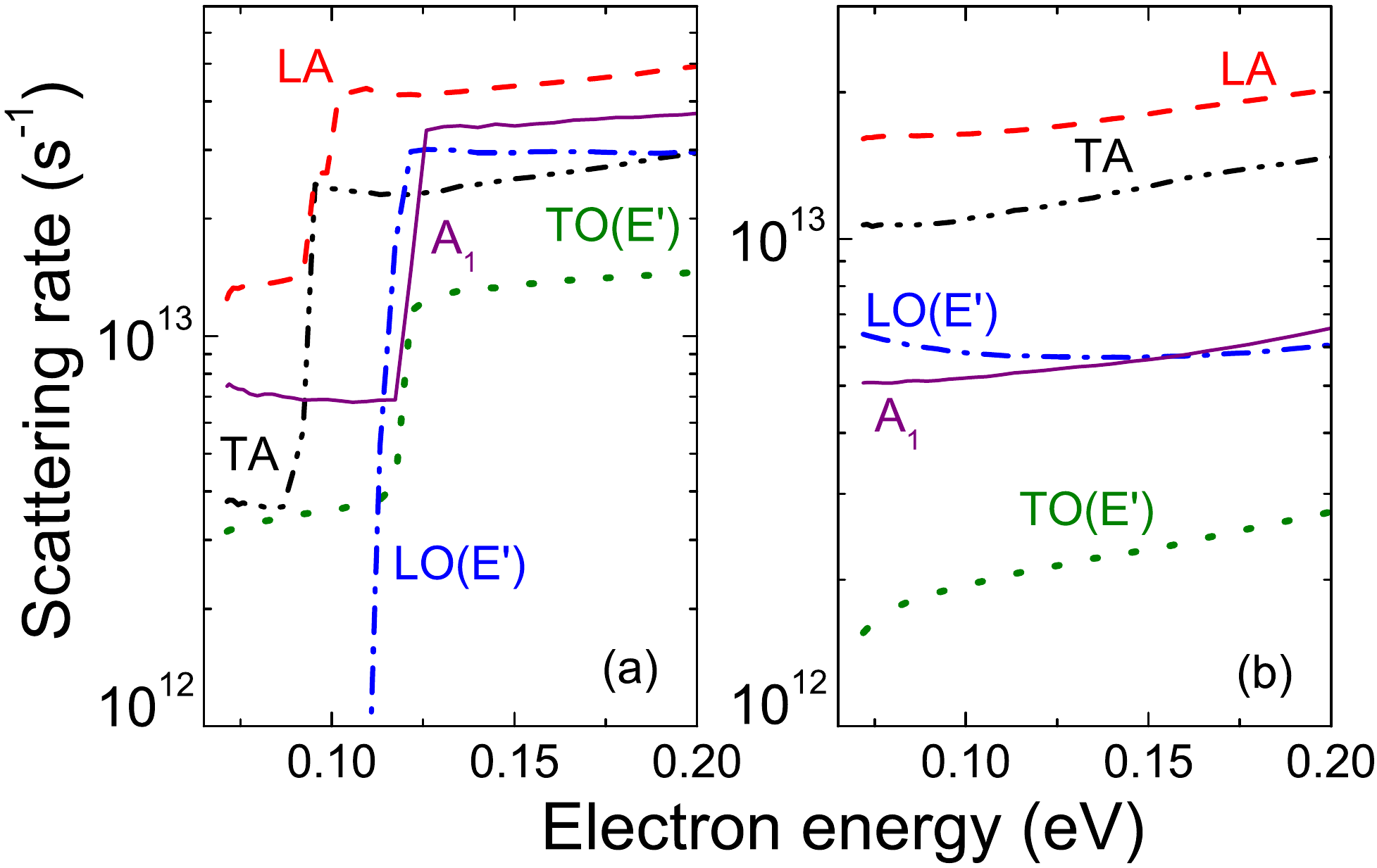}
\caption{ (Color online) Scattering rates of Q-valley electrons in MoS$_2$ via (a) emission and (b) absorption of phonons calculated at room temperature.  The Q-K separation energy $E_\mathrm{QK}$ (= 70 meV) denotes the onset of curves as the K-valley minimum serves as the reference (zero) of energy scale. The electron wavevector $\mathbf{k}$ is assumed to be along the G-$\Gamma$ axis.}
\label{MoS2_SR_Q}
\end{figure}

The electron-phonon scattering rates are calculated as a function of electron energy using Fermi's golden rule.  Figure~\ref{MoS2_SR_K} gives the rates for electrons in the K valleys at room temperature, while the result for Q-valley electrons is shown in Fig.~\ref{MoS2_SR_Q}. Similarly to silicene, the wavevector $\mathbf{k}$ of the initial electronic state  is chosen along the K-$\Gamma$ or Q-$\Gamma$ axis, respectively. As can be seen from the figures, the LA mode provides the largest scattering rates consistent with its large coupling strength. The discontinuities or steps in the curves represent either the onset of optical phonon emission or intervalley scattering.  For instance, the abrupt increase observed in the rate of LA phonons at $\sim$100~meV in Fig.~\ref{MoS2_SR_K}(a) can be attributed to the above mentioned strong $\mathrm{K} \rightarrow \mathrm{Q^{\prime}}$ transition via emission of a LA phonon with $\mathbf{q}=\mathrm{M}$.  Since this phonon energy is approximately 30~meV (see Fig.~\ref{MoS2_band_ph}), the final state energy of $\sim$70~meV indeed matches to the Q-K separation $E_\mathrm{QK}$.  Similarly, the onset of transition via absorption can be found around 40 meV in Fig.~\ref{MoS2_SR_K}(b). The large density of states in the Q valleys (evident from the large effective masses) makes the contribution of this scattering even more prominent.  If, on the other hand, all of the final states in the Q valleys are excluded, the total scattering rate for the K-valley electron reduces drastically to approximately $2\times 10^{13}$~s$^{-1}$, which is consistent with the prediction of an earlier first-principles calculation.~\cite{Kaasbjerg2012} The observed difference of an order of magnitude clearly illustrates the strong dependence of the scattering rates on $E_\mathrm{QK}$.  The inconsistency of this value in the recent publications~\cite{Kadantsev2012,Kaasbjerg2012} adds difficulty to accurately evaluating the role of Q valleys.

\begin{figure}
\centering
\includegraphics[width=8.0cm]{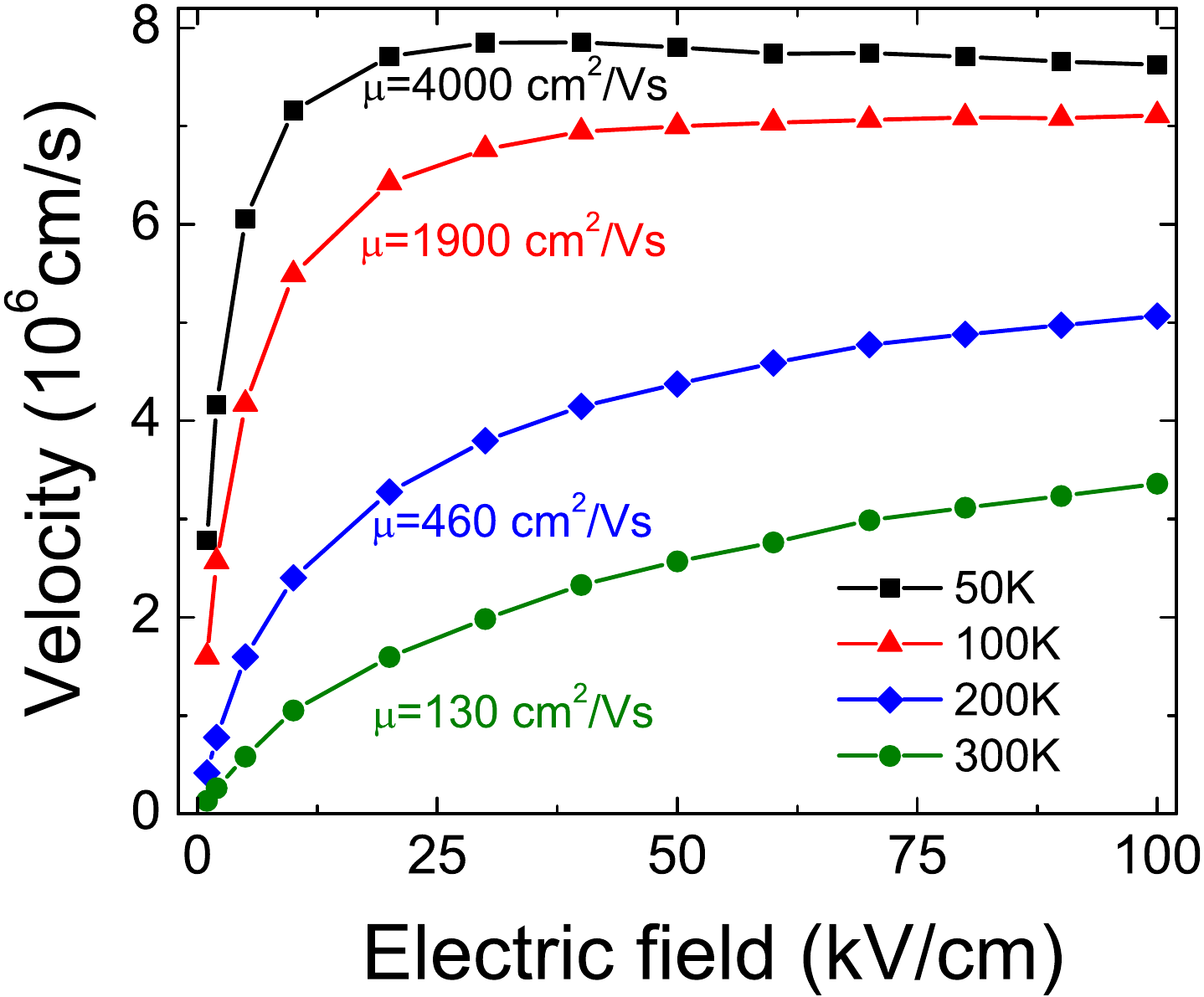}
\caption{(Color online) Drift velocity versus electric field in monolayer MoS$_2$ obtained from a Monte Carlo simulation at different temperatures with $E_\mathrm{QK} = 70$ meV.  When electron transfer to the Q valleys is not considered, the mobility increases to approx. 320 cm$^2$/Vs at 300 K.}
\label{MoS2_velocity}
\end{figure}

Utilizing the scattering rates, the velocity vs.\ field relation is obtained by a full-band Monte Carlo simulation at four different temperatures.  As shown in Fig.~\ref{MoS2_velocity}, the extracted mobility decreases from $4000$~cm$^2$/Vs at 50 K to about $130$~cm$^2$/Vs at room temperature while the saturation velocity changes from $7.6\times 10^6$~cm/s to $3.4\times 10^6$~cm/s. The small mobility and saturation velocity can be attributed to strong electron-phonon scattering as well as the heavy effective masses.  With massive electrons that hinder acceleration and many states to scatter into (e.g., K, K$^\prime$, Q, Q$^\prime$ valleys), this is an expected outcome.

Compared to a recent theoretical estimation~\cite{Kaasbjerg2012} of 410 cm$^2$/Vs and experimentally extracted values~\cite{Lembke2012} as high as 1090 cm$^2$/Vs, however, our mobility is significantly smaller, requiring a careful analysis of the discrepancy.  Two factors are identified that could yield at least a partial explanation. First, let us examine the issues surrounding the Q-K separation. With inconsistencies reported in several first-principles results on this sensitive quantity (see the discussion above), it is not unreasonable to imagine that our DFT calculation may have also experienced similar inaccuracies. If $E_\mathrm{QK}$ proves to be substantially larger than the estimated 70 meV, then the Q valleys would have a limited influence on the low-field mobility and can be neglected in the calculation as in Ref.~\onlinecite{Kaasbjerg2012} (with 200 meV). In this case, our simulation estimates the K-valley dominated mobility of 320 cm$^2$/Vs that is essentially in agreement with the earlier prediction (410 cm$^2$/Vs).~\cite{Kaasbjerg2012}  Clearly, both first-principles models produce a consistent picture of K-valley electron dynamics including intrinsic scattering with relevant phonon modes.  The difference is the relative significance of Q valleys (e.g., 70 meV vs.\ 200 meV). As such, further clarification of intrinsic mobility in monolayer MoS$_2$ may need to be preceded by accurate experimental determination of $E_\mathrm{QK}$.

Even when the influence of Q valleys becomes negligible, a sizable disparity remains between the theoretically obtained mobility and the highest value claimed experimentally~\cite{Lembke2012} (e.g., 320$-$410 cm$^2$/Vs vs.\ 1090 cm$^2$/Vs).  This is puzzling since the theoretical estimates of the intrinsic parameters are generally expected to provide an upper limit to the measured data that tend to experience additional mobility degrading, due to external scattering sources.  When examining Ref.~\onlinecite{Lembke2012}, however, it becomes clear that the extracted data is not a direct measurement of intrinsic mobility; rather, it is the channel mobility that is strongly influenced by the details of the structure and the bias conditions. A particularly salient point is that the deduction relied on the transistor I-V data measured when the MoS$_2$ channel is populated  by electrons (i.e., degenerate). This condition deviates from the underlying assumption of nondegeneracy in our calculation, necessitating consideration of, among others, the screening effect.
In a low-dimensional system, it is known that the screening due to the degenerate electrons can lower the bare scattering rates substantially (thus, affecting the mobility),~\cite{Kawamura1992,Tanatar1993} as previously demonstrated experimentally in the AlGaAs/GaAs structures.~\cite{Henriksen2005} The screening effect can be included by renormalizing the electron-phonon interaction through the dielectric function; i.e., $\textsf{g}_{\mathbf{q},\mathbf{k}}^{v}
\rightarrow\textsf{g}_{\mathbf{q},\mathbf{k}}^{v}/{\epsilon(q)}$.~\cite{Kawamura1992,Tanatar1993}

To gauge the potential significance in monolayer MoS$_2$, we adopt a simple model for the dielectric function based on Thomas-Fermi screening of only the K-valley electrons: i.e., $\epsilon(q)=1+{q_\mathrm{TF}}/{q}$, where the screening wavevector $q_\mathrm{TF}={4m_\mathrm{K}e^2}/{\hbar^2\kappa}$. Here, the factor of $4$ accounts for the spin and valley degeneracies, $e$ is the electron charge, and $\kappa$ is the background dielectric constant.
Subsequent calculation with a rough estimate of $\epsilon(q)$ shows that the scattering rates can experience a decrease of about an order of magnitude through screening. A corresponding increase of the mobility is estimated to be well over 1000~cm$^2$/Vs that is more consistent with the value extracted from the transistor I-V characteristics.~\cite{Lembke2012}  Thus, it is evident that the screening must be taken into account accurately when the carrier density becomes degenerate in MoS$_2$.  A detailed analysis of $\epsilon(q)$ is, however, beyond the scope of this investigation as our focus is on the properties of intrinsic electron-phonon interaction.



\subsection{Deformation Potential Model}
For practical applications, it would be convenient to approximate the
\emph{ab initio} results for electron-phonon coupling by a simple
analytical model.  Particularly useful in the present case is the deformation potential approximation.  Under this treatment, the coupling matrix $\langle j,\mathbf{k}+\mathbf{q}\vert\bigtriangleup V_{\mathbf{q},\mathrm{SCF}}^{v}\vert i,\mathbf{k}\rangle$ in Eq.~(\ref{matrix}) can be expressed in the first order ($D_{1}\mathbf{q}$), or in the zeroth order ($D_{0}$).~\cite{Ferry2000}
The first-order deformation potential constant ($D_1$) is adopted to represent the coupling matrices for the acoustic phonon modes in the long wavelength limit (i.e., intravalley scattering).
In comparison, those involving the near zone-edge acoustic phonons (i.e., intervalley scattering) are treated by using the zero$^{th}$-order deformation potential ($D_0$) in a manner analogous to the optical modes. In the latter case ($D_0$), the phonon energy is assumed independent of the momentum for simplicity.  The obtained analytical expressions of the scattering rates are then matched to the first-principles results by fitting the effective deformation potential constants.

For silicene, the {\em intravalley} scattering rate by acoustic mode $v$ (= LA, TA, ZA) is obtained as
\begin{equation}\label{Dap1_Silicene}
\left. \frac{1}{\tau_{\mathbf{k},v}^{(1)}} \right\vert_{\mathrm{Si}} =\frac{D_{1}^2k_BT}{\hbar^3 v_{F}^2\rho v_{v}^2}E_{\mathbf{k}} \,.
\end{equation}
Here $\rho$ is the mass density ($= 7.2\times 10^{-8}$ g/cm$^2$) and $v_{v}$ denotes the sound velocity, for which we can take the value of $v_\mathrm{ZA}=6.3\times 10^4$~cm/s, $v_\mathrm{TA}=5.4\times 10^5$~cm/s, and $v_\mathrm{LA}=8.8\times 10^5$~cm/s, respectively, as discussed earlier. On the other hand, the rate of optical phonon scattering (both {\em intravalley} and {\em intervalley} transitions) as well as the {\em intervalley} acoustic phonon scattering can be expressed by the following form:
\begin{equation}\label{Dop0_Silicene}
\left. \frac{1}{\tau_{\mathbf{k},v}^{(2)}} \right\vert_{\mathrm{Si}} =\frac{D_{0}^2}{2\hbar^2 v_F^2\rho \omega_{v}} [(E_{\mathbf{k}}+\hbar\omega_{v})N_v+(E_{\mathbf{k}}- \hbar\omega_{v})(N_v+1) \Theta(E_{\mathbf{k}}-\hbar\omega_{v})],
\end{equation}
where $\Theta$(x) is the Heavyside step function and $N_{v}=[\exp({\hbar\omega_{v}}/{k_B T})+1]^{-1}$ for phonon mode $v$.  Since the phonon dispersion is treated constant in Eq.~(\ref{Dop0_Silicene}), the {\em intravalley} optical phonon scattering approximates the zone-center (i.e., $\Gamma$) phonon energies for $\hbar\omega_{v}$ ($v =$ LO, TO, ZO).  In the case of {\em intervalley} scattering via acoustic or optical phonons, $\hbar\omega_{v}$ takes the respective phonon energy at the zone-edge K point corresponding to electron transition $\mathrm{K} \leftrightarrow \mathrm{K^\prime}$.  The specific values used in the calculation can be found in Table~\ref{table_ph_Silicene}.   When matched to the first-principles rates, the effective deformation potential constants can be extracted for each scattering process as summarized in Table~\ref{table_Silicene}.  A particularly interesting point to note from the result is that all three acoustic phonons show generally comparable values of $D_1$ and $D_0$ despite the large scattering rate of ZA mode (see also Fig.~\ref{Silicene_SR}).  Clearly, this mode (ZA) couples strongly with electrons but not enough to prevail over other acoustic branches; its dominant contribution is due mainly to the small phonon energy as discussed earlier.

The electron-phonon scattering processes in MoS$_2$ are much more complicated as the deformation potentials need to be determined for both K- and Q-valley electrons. While feasible, it is not practically useful to define the effective interaction constants individually based on the mode of involved phonons and the transition types.  Accordingly, we adopt a simplified description by combining the appropriate contributions into just two modes, acoustic and optical, respectively.

Using the effective mass approximation for the band structure near the valley minima, the scattering rate for {\em intravalley} acoustic phonon scattering (i.e., $\mathrm{K} \rightarrow \mathrm{K}$ or $\mathrm{Q_1} \rightarrow \mathrm{Q_1}$ by both LA and TA phonons; see Figs.~\ref{MoS2_Matrix} and \ref{MoS2_MatrixQ}) is given by
\begin{equation}\label{Dap1_MoS2}
\left. \frac{1}{\tau_{\mathbf{k}}^{(1)}} \right\vert_{\mathrm{MoS_2}} =\frac{m_p^*D_{1}^2k_BT}{\hbar^3\rho v_{s}^2},
\end{equation}
where $\rho =3.1\times 10^{-7}$ g/cm$^2$ for MoS$_2$ and $m_p^*$ is the density-of-states effective mass for the K or Q valley (final state), $m_p^*=\sqrt{m_p^l m_p^t}$ ($p=$ K,Q). For electrons in the Q valleys, strictly speaking, an additional factor of $\Theta(E_{\mathbf{k}}-E_\mathrm{QK})$ is multiplied to the left side of Eq.~(\ref{Dap1_MoS2}) to account for the non-zero energy at the bottom of the Q valley. By taking the sound velocity $v_s = v_{LA}$ ($=6.6\times 10^5$ cm/s), the value of $D_1$ is estimated to be 4.5 eV and 2.8 eV in the K and Q valleys, respectively.

The analytical expression that describes {\em intravalley} and {\em intervalley} optical phonon scattering as well as {\em intervalley} acoustic phonon scattering rate is obtained as
\begin{equation}\label{Dop0_MoS2}
\left. \frac{1}{\tau_{\mathbf{k},v}^{(2)}} \right\vert_{\mathrm{MoS_2}} =g_d\frac{m_p^*D_{0}^2}{2\hbar^2\rho\omega_{v}}[N_v\triangle_1
+(N_v+1)\triangle_2],
\end{equation}
where $g_d$ is the valley degeneracy for the final electron states, and $\triangle_1$ and $\triangle_2$ denote the onset of scattering for phonon absorption and emission, respectively. For instance, $\triangle_1=1$ and $\triangle_2=\Theta(E_{\mathbf{k}}-\hbar\omega_{v})$ for electron transitions between the K valleys, whereas $\triangle_1=\Theta(E_{\mathbf{k}}-E_\mathrm{QK})$ and $\triangle_2=\Theta(E_{\mathbf{k}}-\hbar\omega_{v}-E_\mathrm{QK})$ for transitions between the Q valleys.  The factors corresponding to intervalley transfer between K and Q valleys can be constructed accordingly, where $E_\mathrm{QK}$ may be treated as a potentially adjustable parameter.  Tables~\ref{table_MoS2_K} and \ref{table_MoS2_Q} summarize the initial/final electron states, the phonon momentum that is involved (in the form of its location in the momentum space), and the extracted deformation potential constants for each transition process considered in the investigation.  For a given phonon momentum, the actual value $\hbar \omega_v$ used in the analytical calculation is the average of the relevant phonon modes.  Specifically, the acoustic (optical) phonon energy is obtained as the average of LA and TA [TO(E$^\prime$), LO(E$^\prime$), and A$_1$] modes at the respective symmetry points given in Table~\ref{table_ph_MoS2}.  An additional point to note is that the estimate of $D_0^{op}$ at the ${\Gamma}$ point includes the effect of Fr\"{o}hlich scattering by the  LO(E$^\prime$) mode.~\cite{Kaasbjerg2012}  While this is a mechanism physically distinct from the deformation potential interaction and must be handled separately, its impact is relatively modest, at least for the electrons in the K valley.  Accordingly, the present treatment is considered adequate.  Further simplification of the model may also be possible judging from the narrow range of values in $D_0^{ac}$ and $D_0^{op}$ (mostly in the low to mid 10$^8$~eV/cm).

\section{Summary}

We have performed a first-principles calculation together with a full-band Monte Carlo analysis to examine  electron-phonon interaction and the intrinsic transport properties in monolayer silicene and MoS$_2$.  The results clearly elucidate the role of different branches as well as the intra/inter-valley scattering.
The predicted intrinsic mobility for silicene is approximately 1200 cm$^2$/Vs, with saturation velocity of $3.9 \times 10^6$~cm/s at room temperature.  In the case of MoS$_2$, the K-valley dominated mobility gives approximately 320 cm$^2$/Vs, while the intrinsic value reduces to about 130 cm$^2$/Vs when the energy separation of 70 meV is used between the K and Q minima.  The estimated saturation velocity is $3.7\times 10^6$~cm/s.  The investigation also illustrates the significance of extrinsic screening, particularly in numerical evaluation of transport characteristics.  The extracted deformation potential constants may prove to be useful in further studies of these materials.

\begin{acknowledgments}
This work was supported, in part, by SRC/FCRP FAME, SRC/NRI SWAN as well as SRC
CEMPI at the University of North Texas (Task ID P14924).  MBN also wishes to acknowledge partial support from the Office of Basic Energy Sciences, US DOE at Oak Ridge National Lab under contract DE-AC05-00OR22725 with UT-Battelle, LLC.  Calculations have been run at NCCS-ORNL and the NCSU-HPC Initiative.

\end{acknowledgments}

\clearpage

\begin{table}[ht]
\caption{Phonon energies (in units of meV) at the symmetry points for monolayer silicene.}
\begin{tabular}{c c c c}
\hline\hline
Phonon modes~~&~~~$\mathrm{\Gamma}$~~~&~~~$\mathrm{K}$~~~&~~~$\mathrm{M}$~~~   \\ [-0.5ex]
\hline
ZA & 0 & 13.2 & 13.0  \\
TA & 0 & 23.7 & 13.4  \\
LA & 0 & 13.2 & 13.5  \\
ZO & 22.7 & 50.6 & 50.7 \\
TO & 68.8 & 50.6 & 56.7 \\
LO & 68.8 & 61.7 & 64.4 \\
\hline
\label{table_ph_Silicene}
\end{tabular}
\end{table}

\clearpage

\begin{table}[ht]
\caption{Phonon energies (in units of meV) for TA, LA, TO(E$^\prime$), LO(E$^\prime$), and A$_1$ (or homopolar) modes at the $\mathrm{\Gamma}$, $\mathrm{K}$, $\mathrm{M}$ and $\mathrm{Q}$ points in the FBZ of monolayer MoS$_2$.}
\begin{tabular}{c c c c c}
\hline\hline
Phonon modes~~&~~~$\mathrm{\Gamma}$~~~&~~~$\mathrm{K}$~~~&~~~$\mathrm{M}$~~~&~~~$\mathrm{Q}$~~~ \\ [-0.5ex]
\hline
TA & 0 & 23.1 & 19.2 & 17.9 \\
LA & 0 & 29.1 & 29.2 & 23.6 \\
TO(E$^\prime$) & 48.6 & 46.4 & 48.2 & 48.0 \\
LO(E$^\prime$) & 48.9 & 42.2 & 44.3 & 44.2 \\
A$_1$ & 50.9 & 51.9 & 50.1 & 52.2 \\
\hline
\label{table_ph_MoS2}
\end{tabular}

\end{table}

\clearpage

\begin{table}[ht]
\caption{Extracted deformation potential constants for electron-phonon interaction in  silicene.}
\centering
\begin{tabular}{c c c}
\hline\hline
Phonon mode~~&~~~~~Intravalley~~~~~&~~~~~Intervalley~~~~~ \\ [0.5ex]
\hline
ZA  &  2.0 eV & 6.1$\times$10$^7$~eV/cm  \\
TA  &  8.7 eV & 1.4$\times$10$^8$~eV/cm  \\
LA  &  3.2 eV & 4.2$\times$10$^7$~eV/cm  \\
ZO  &  6.3$\times$10$^7$~eV/cm & 4.3$\times$10$^7$~eV/cm \\
TO  &  1.8$\times$10$^8$~eV/cm & 1.4$\times$10$^8$~eV/cm \\
LO  &  1.9$\times$10$^8$~eV/cm & 1.7$\times$10$^8$~eV/cm \\
\hline
\label{table_Silicene}
\end{tabular}

\end{table}

\clearpage
\begin{table}[ht]
\caption{Extracted deformation potential constants for electron-phonon interaction in MoS$_2$ for electrons in the K valley [see also Fig.~\ref{MoS2_Matrix}(f)].}
\centering
\begin{tabular}{c c c}
\hline\hline
Phonon &  ~~~Electron~~~ & Deformation   \\ [-0.5ex]
Momentum &  ~~~Transition~~~  & potentials \\
\hline
 &  & $D_{1}^{ac}$=4.5 eV   \\
\raisebox{1.5ex}{${\Gamma}$}&  \raisebox{1.5ex}{$\mathrm{K}\rightarrow\mathrm{K}$} &  $D_{0}^{op}$=$5.8\times 10^8$~eV/cm \\
\hline

&  &  $D_{0}^{ac}$=$1.4\times 10^8$~eV/cm \\
\raisebox{1.5ex}{$\mathrm{K^\prime}$} & \raisebox{1.5ex}{$\mathrm{K}\rightarrow\mathrm{K^\prime}$} &$D_{0}^{op}$=$2.0 \times 10^8$~eV/cm \\
\hline

 &  & $D_{0}^{ac}$=$9.3 \times 10^7$~eV/cm   \\
\raisebox{1.5ex}{$\mathrm{Q^\prime}$} &\raisebox{1.5ex}{$\mathrm{K}\rightarrow\mathrm{Q}$}  & $D_{0}^{op}$=$1.9\times 10^8$~eV/cm \\
\hline

 & &  $D_{0}^{ac}$=$4.4\times 10^8$~eV/cm \\
\raisebox{1.5ex}{$\mathrm{M}$}  &\raisebox{1.5ex}{$\mathrm{K}\rightarrow\mathrm{Q^\prime}$} &    $D_{0}^{op}$=$5.6 \times 10^8$~eV/cm  \\
\hline
\label{table_MoS2_K}
\end{tabular}

\end{table}

\clearpage

\begin{table}[ht]
\caption{Extracted deformation potential constants for electron-phonon interaction in MoS$_2$ for  electrons in the $\mathrm{Q_1}$ valley [see also Fig.~\ref{MoS2_MatrixQ}(f)].  Multiple equivalent valleys for the final state specify the degeneracy factor $g_d$ larger than one in Eq.~(\ref{Dop0_MoS2}). }
\centering
\begin{tabular}{c c c c}
\hline\hline
Phonon &  ~~~Electron~~~ & Deformation   \\ [-0.5ex]
Momentum &  ~~~Transition~~~ & potentials \\
\hline
 &  & $D_{1}^{ac}$=2.8 eV   \\
\raisebox{1.5ex}{$\mathrm{\Gamma}$}&  \raisebox{1.5ex}{$\mathrm{Q_1}\rightarrow\mathrm{Q_1}$}  & $D_{0}^{op}$=$7.1 \times 10^8$~eV/cm \\
\hline

&  &  $D_{0}^{ac}$=$2.1 \times 10^8$~eV/cm \\
\raisebox{1.5ex}{$\mathrm{Q_3} (\mathrm{Q_5})$} & \raisebox{1.5ex}{$\mathrm{Q_1}\rightarrow\mathrm{Q_2} (\mathrm{Q_6})$}  &$D_{0}^{op}$=$4.8\times10^8$~eV/cm \\
\hline
 &  & $D_{0}^{ac}$=$2.0\times 10^8$~eV/cm   \\
\raisebox{1.5ex}{$\mathrm{M_3} (\mathrm{M_4})$} &\raisebox{1.5ex}{$\mathrm{Q_1}\rightarrow\mathrm{Q_3} (\mathrm{Q_5})$} & $D_{0}^{op}$=$4.0\times10^8$~eV/cm \\
\hline
 & & $D_{0}^{ac}$=$4.8\times10^8$~eV/cm \\
\raisebox{1.5ex}{$\mathrm{K^\prime}$}  &\raisebox{1.5ex}{$\mathrm{Q_1}\rightarrow\mathrm{Q_4}$} &    $D_{0}^{op}$=$6.5\times 10^8$~eV/cm  \\
\hline
 & & $D_{0}^{ac}$=$1.5\times10^8$~eV/cm \\
\raisebox{1.5ex}{$\mathrm{Q_1}$}  &\raisebox{1.5ex}{$\mathrm{Q_1}\rightarrow\mathrm{K}$} &    $D_{0}^{op}$=$2.4\times10^8$~eV/cm  \\
\hline
 & & $D_{0}^{ac}$=$1.5\times 10^8$~eV/cm \\
\raisebox{1.5ex}{$\mathrm{M_2} (\mathrm{M_5})$}  &\raisebox{1.5ex}{$\mathrm{Q_1}\rightarrow\mathrm{K^\prime}$} &  $D_{0}^{op}$=$2.4\times10^8$~eV/cm  \\

\hline
\label{table_MoS2_Q}
\end{tabular}

\end{table}

\clearpage

\clearpage

\clearpage

\clearpage

\clearpage

\clearpage

\clearpage

\clearpage

\clearpage

\clearpage

\end{document}